\documentclass{article}

\usepackage[final, nonatbib]{neurips_2021}

\usepackage[utf8]{inputenc} 
\usepackage[T1]{fontenc}    
\usepackage{hyperref}
\usepackage{academicons}
\usepackage{xcolor}
\newcommand{\orcid}[1]{\href{https://orcid.org/#1}{\textcolor[HTML]{A6CE39}{\aiOrcid}}}
\usepackage{url}            
\usepackage{booktabs}       
\usepackage{amsfonts}       
\usepackage{nicefrac}       
\usepackage{microtype}      
\usepackage{xcolor}         
\usepackage{graphicx}
\usepackage{dcolumn}
\usepackage{bm}
\usepackage{tikz}
\usetikzlibrary{svg.path}

\definecolor{orcidlogocol}{HTML}{A6CE39}
\tikzset{
  orcidlogo/.pic={
    \fill[orcidlogocol] svg{M256,128c0,70.7-57.3,128-128,128C57.3,256,0,198.7,0,128C0,57.3,57.3,0,128,0C198.7,0,256,57.3,256,128z};
    \fill[white] svg{M86.3,186.2H70.9V79.1h15.4v48.4V186.2z}
                 svg{M108.9,79.1h41.6c39.6,0,57,28.3,57,53.6c0,27.5-21.5,53.6-56.8,53.6h-41.8V79.1z M124.3,172.4h24.5c34.9,0,42.9-26.5,42.9-39.7c0-21.5-13.7-39.7-43.7-39.7h-23.7V172.4z}
                 svg{M88.7,56.8c0,5.5-4.5,10.1-10.1,10.1c-5.6,0-10.1-4.6-10.1-10.1c0-5.6,4.5-10.1,10.1-10.1C84.2,46.7,88.7,51.3,88.7,56.8z};
  }
}

\newcommand\orcidicon[1]{\href{https://orcid.org/#1}{
\begin{tikzpicture}[xscale=0.05,yscale=0.05,transform shape]
\pic{orcidlogo};
\end{tikzpicture}
}}
\title{Constraining cosmological parameters from N-body simulations with Bayesian Neural Networks}

\author{
      H\'ector J. Hort\'ua \orcidicon{0000-0002-3396-2404}  \\
   Semillero de investigaci\'on en Data Science, \\Escuela de Ciencias Básicas Tecnologías e Ingenierías ECBTI,\\
   Universidad Nacional Abierta y a Distancia UNAD\\
   CEAD José Acevedo y Gómez - Bogotá- Colombia \\
  \texttt{hector.hortua@unad.edu.co} }

\begin{document}

\maketitle

\begin{abstract}
  In this paper we use   \texttt{The Quijote} simulations in order to extract the cosmological parameters through Bayesian Neural Networks. This kind of models has a  remarkable ability of estimating the associated uncertainty, which is one of the ultimate goals in the precision cosmology era.
  We demonstrate the advantages of   BNNs for extracting  more complex output distributions and  non-Gaussianities information from the simulations. 
\end{abstract}

\section{Introduction}

Cosmological N-body simulations provide a unique means to test gravity on large-scales. It opens the possibility to fully explore the growth of structure in both the linear and non-linear regime \cite{simul}. Currently, the concordance cosmological model, $\Lambda$-Cold Dark Matter ($\Lambda$CDM), gives an accurate description of most the observations from early universe as well as late stages using a set of six parameters. The most recent observations from Cosmic Microwave Background (CMB) and cosmic distance ladder estimations have shown a tension between some of the parameters: $\sigma_8$ being the amplitude of the power spectrum normalization and $h$ being the normalized Hubble constant. At the same time exists a well known degeneracy with the total non-relativistic matter density parameter $\Omega_m$ \cite{Tinker}. 

Recently, Neural networks have been used on a variety of tasks because of their potential to  solving under determined inverse problems.    However,  neural networks are prone to overfitting due to the excessive number of parameters to be adjusted and do not provide a measure of uncertainty for its predictions.  This problem can be addressed by following a probabilistic approach that permit quantifying the uncertainty of the model predictions \cite{Chang,Hortua}. In order to do that, \texttt{The Quijote simulations} \cite{Villaescusa} are used to extract cosmological information via two approaches: a traditional neural network and a Bayesian one, which has the advantage of estimating the associated uncertainty. The main contributions of this paper are the following:

\begin{enumerate}
    \item Propose probabilistic neural networks approaches for cosmological analysis and provide  alternative techniques for extracting complex information from simulations.
    \item Compare the performance and advantages of BNNs with respect to the traditional architectures in physics.
    
\end{enumerate}

\section{Dataset and Network}

In this work, we  leverage  2000  latin-hypercubes simulation taken from \texttt{The Quijote} project \cite{Villaescusa}.  These simulations  vary the value of each parameter as following: $\Omega_m \in [0.1, 0.5]$, $h \in [0.6, 0.9]$ and $\sigma_8 \in [0.6, 1]$. To account for the presence of cosmic variance, the simulations were running with different random seeds and high-resolution \cite{Villaescusa}. We make use of $32^3$- and $64^3$-volxes as inputs for the Neural Net (interpolated from the standard resolution simulations $512^3$ points) that contain information of the three-dimensional density field. The output of the network consists in nine neurons: three  for the mean of the three parameters, and six  for the elements of the covariant  matrix. Those outputs fed the next layer which is  a Multivariate Gaussian distribution. 
TensorFlow and TensorFlow-probability\footnote{https://www.tensorflow.org/} were used for building the traditional and stochastic Neural Networks. We have used the Mean Square Error (MSE) as loss function  for the deterministic network, while for BNN the minimization is applied  to the negative-log-likelihood of the distribution on the top of the network. In both scenarios, the coefficient of determination ($R^2$) was used to assess the performance:
\begin{equation}
  R^2=1-\frac{\sum_i (\bm{\bar{\mu}}(\bm{x}_i)-\bm{y}_i)^2}{\sum_i (\bm{y}_i-\bm{\bar{y}})^2},
\end{equation}

where $\bm{\bar{\mu}}(\bm{x}_i)$ are the predicted values of the trained Network, $\bm{\bar{y}}$ is the average of the true parameters and the summations are performed over the entire test set. $R^2$ ranges from 0 to 1, where 1 represents perfect inference.

\section{Method}

The  architecture employed in this work is shown in Fig.\ref{fig3} where the weights were updated via backpropagation. Then, after obtaining a decent performance in the model, we froze the entire  network and put on the top  a dense layer which leverages the  Flipout estimator to make a Monte Carlo approximation of the distribution in the kernel and bias \cite{Mozer}.   BNNs are useful for obtaining the aleatoric uncertainties that represent the intrinsic randomness in the input dataset  and they can be reduced enhancing the quality of the data \cite{Hortua} . On the other hand, epistemic uncertainty quantifies the ignorance about the correct model that generated the data, it includes the uncertainty in the model, parameters, and convergence. This uncertainty is caused by the limited training data with respect to the entire feature space. Collecting more data in regions where there is a low density of training examples will reduce this uncertainty, while the aleatoric will remain
unchanged. Methods for estimating epistemic uncertainties are different from the aleatoric ones, and this is where BNNs can offer a mathematically grounded base for computing this uncertainty and be able to estimate the performance of the model  \cite{Hortua}.

\section{Results}
\label{results}
\begin{figure}[h!]
\begin{center}
\includegraphics[width=1.01\textwidth]{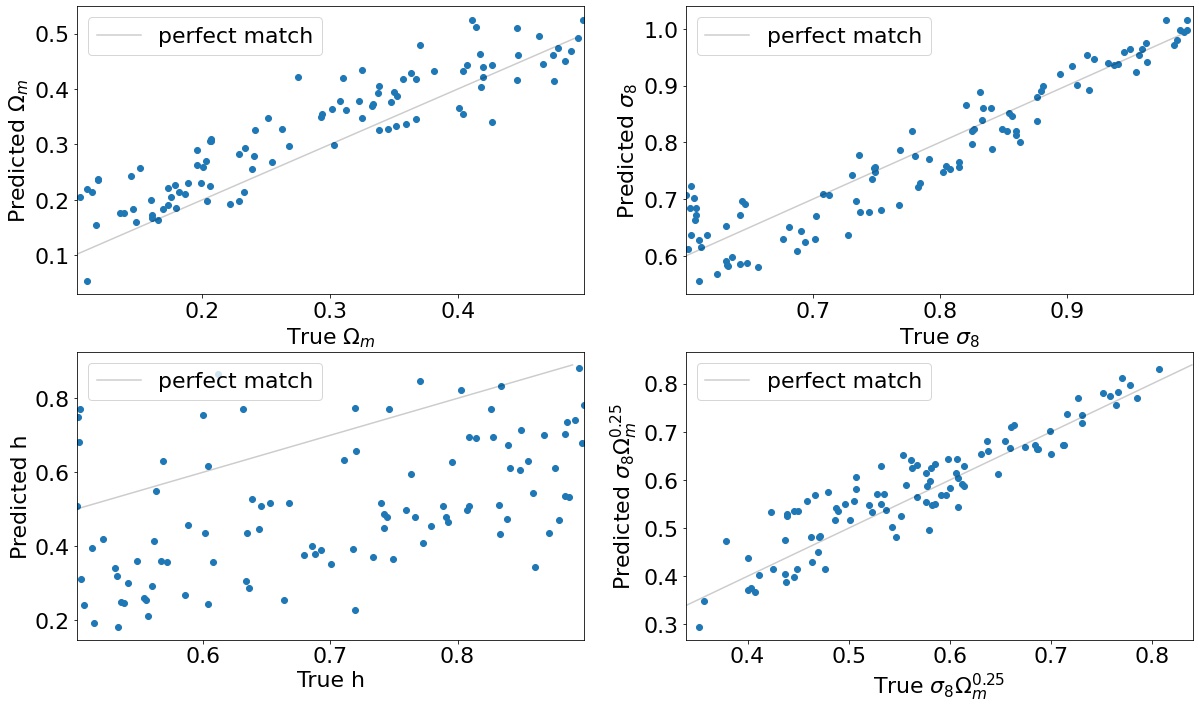}
\end{center}
\caption{\it Plot of predicted values of three parameters over the target values of the convolutional neural network.  The solid blue line represents a perfect predictor.} \label{fig1}
\end{figure}

\begin{figure}[h!]
\begin{center}
\includegraphics[width=0.9\textwidth]{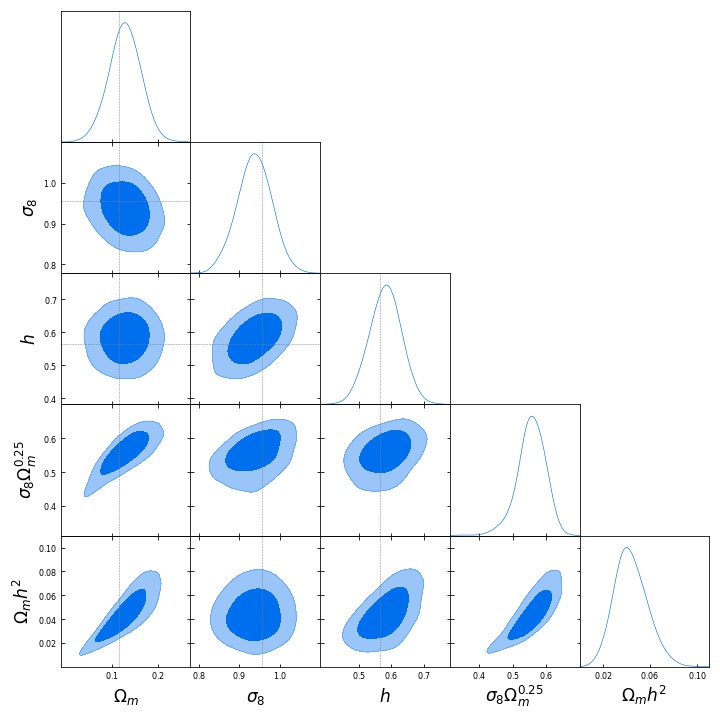}
\end{center}
\caption{\it Constraints on the value of the cosmological parameters from the simulations. The light-blue and dark-blue regions represent the $1\sigma$ and $2\sigma$ constraints, respectively. The panels with
the solid lines represent the true values.} \label{fig2}
\end{figure}

Fig.\ref{fig1} reports the prediction  versus the true values for  samples. As we observed, the model can capture most of the information encoded in the dataset. The performance can be also seen in Table\ref{sample-table}, where we observed that $\sigma_8$ is predicted with high accuracy  as it was also reported in \cite{Lazanu}.  On the other hand, the Bayesian Neural Networks offer better probabilistic inference estimates, in terms of their consideration of uncertainty. The results of this model are displayed in Fig.\ref{fig2}. Here, a sample from the test set was taken in order to visualize the parameter constraints. Correlations among parameters are displayed in the triangular plot, which permit to analyze the features of the parameters and describe cosmological probes which allow to  break degeneracies. Additionally, during inference, the model provides faster samples with respect to classical methods like MCMC which can  bring significant advantages in huge datasets.  Normalizing flows and other architectures like Resnet-3D are other ingredients used during this preliminary research which will be described in more detail in a future report.

\begin{table}[h!]
  \caption{Best performance described by coefficient of determination $R^2$
reached for convolutional and Bayesian neural network; along with the constrains for the parameters at $95\%$ C.L. predicted by the BNN. True values correspond to the real values for the sample used for the BNN test visualization. }
  \label{sample-table}
  \centering
  \begin{tabular} { l|  c c c|| c}

 Parameter &  95\% limits &True value& $R^2$ BNN &$R^2$ CNN \\
\hline
{\boldmath$\Omega_m       $} & $0.129^{+0.067}_{-0.068}  $&0.115 &0.833& 0.871 \\

{\boldmath$\sigma_8       $} & $0.935^{+0.084}_{-0.084}   $&0.955&0.881&0.915\\

{\boldmath$h              $} & $0.582^{+0.098}_{-0.097}   $&0.564&0.401&0.456\\

$\sigma_8\Omega_m^{0.25}   $ & $0.555^{+0.080}_{-0.094}   $&0.557&--&--\\

$\Omega_m h^{2}            $ & $0.044^{+0.029}_{-0.027}   $&0.368&--&--\\
\hline
\end{tabular}
\end{table}
\section{Discussion}
\label{headings}
In cosmology, the large-scale, three-dimensional matter distribution, modeled with N-body simulations, plays an important role in understanding the evolution of the universe.
In this work we have shown the use of  deep convolutional  and Bayesian neural network to estimate cosmological parameters from those N-body simulations.
  Finally, other architectures like ResNet-3D was also used to carry out this task along with  masked autoregressive flows. The results driven by those additional ingredients will be shown in a future publication.


{
\small

}


\appendix

\section{Appendix}
\begin{figure}[h!]
\begin{center}
\includegraphics[width=0.7\textwidth]{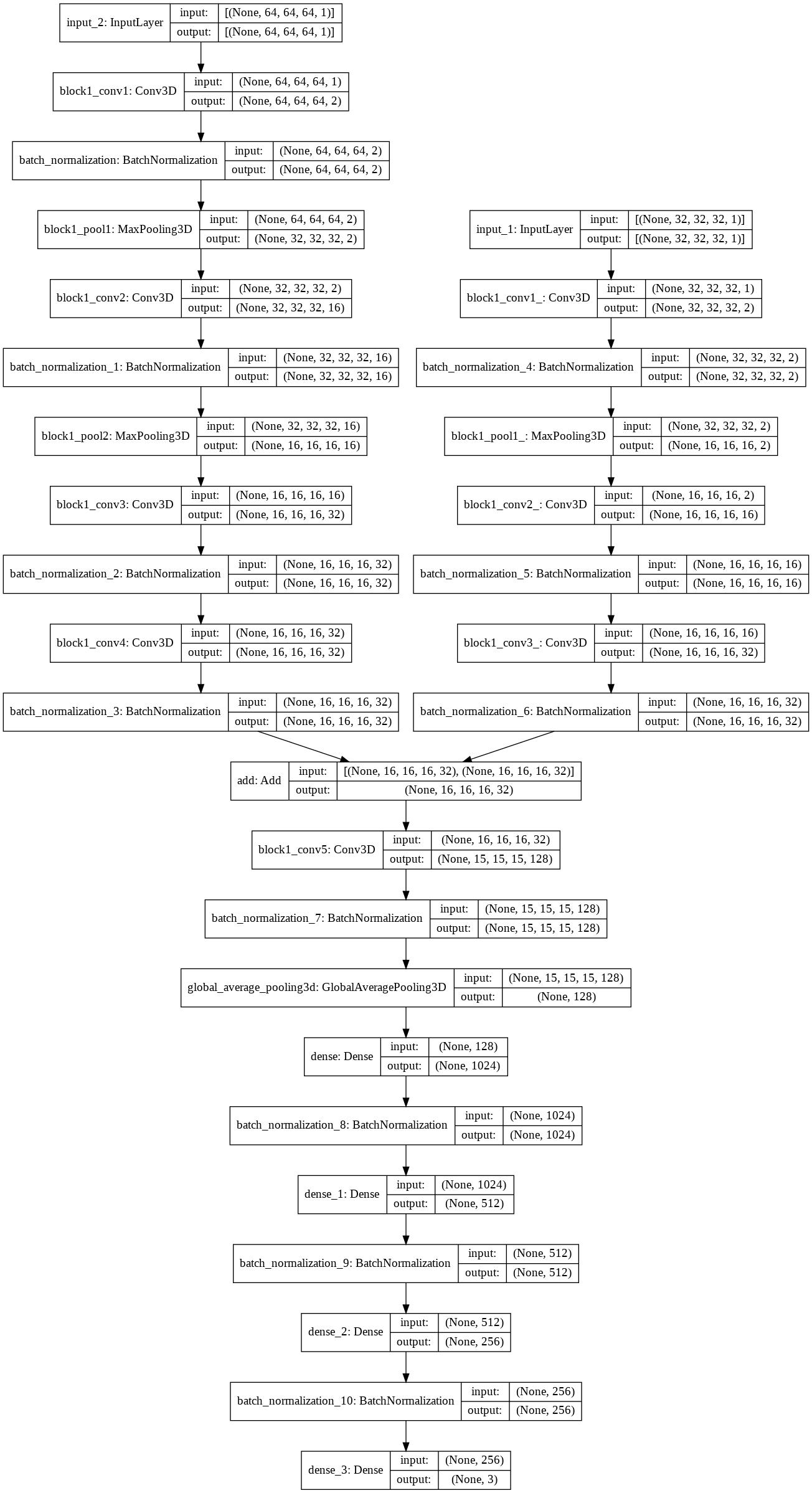}
\end{center}
\caption{\it CNN architecture used in the paper. This is an multi-input model with $32^3$- and $64^3$-volxes. } \label{fig3}
\end{figure}

Fig.\ref{fig3} shows the architecture used in this paper and also, works as backbone for the BNN. 

\end{document}